\documentclass[12pt]{article}
\usepackage{amsmath}
\usepackage{epsfig}
\usepackage{verbatim}
\begin{document}
%
%
\begin{center}
\LARGE
\textbf{Can Everett be Interpreted Without
Extravaganza?}\\[1cm]
\large
\textbf{Louis Marchildon}\\[0.5cm]
\normalsize
D\'{e}partement de physique,
Universit\'{e} du Qu\'{e}bec,\\
Trois-Rivi\`{e}res, Qc.\ Canada G9A 5H7\\
email: marchild$\hspace{0.3em}a\hspace{-0.8em}
\bigcirc$uqtr.ca\\
\end{center}
\medskip
%
%
%
%
\begin{abstract}
Everett's relative states interpretation of
quantum mechanics has met with problems related
to probability, the preferred basis, and
multiplicity.  The third theme, I argue, is the most
important one.  It has led to developments of the
original approach into many-worlds, many-minds,
and decoherence-based approaches.
The latter especially have been advocated in recent
years, in an effort to understand multiplicity
without resorting to what is often perceived
as extravagant constructions.  Drawing from and
adding to arguments of others,
I show that proponents of decoherence-based
approaches have not yet succeeded in making
their ontology clear.
\end{abstract}
%
\section{Introduction}
\label{sec1}
Everett's `relative states' formulation of quantum
mechanics was proposed more than 50 years
ago~\cite{everett1,everett2}, at a time when the
Copenhagen interpretation reigned essentially
unchallenged.  Everett wanted to (i) retain
the universal validity of the Schr\"{o}dinger
equation; (ii) eliminate the need for the
collapse of the wave function; (iii) eliminate
the need for an external observer; and (iv)
offer a derivation of the Born rule.

Research on Everett's approach has, over the
years, largely focussed on the three themes of
probability, the preferred basis, and multiplicity.
This paper intends to succinctly assess where this
research stands and what are the most significant
open problems.  I shall review the above three themes,
emphasizing why I believe the third one is the most
important.  I will then comment on different ways of
understanding multiplicity, arguing that the
currently most popular one leaves crucial questions
unanswered.
%
\section{Probability and preferred basis}
\label{sec2}
Everett claimed that ``the statistical assertions
of the usual interpretation \mbox{[\ldots]} are deducible
(in the present sense) from the pure wave mechanics
that starts completely free of statistical
postulates''~\cite{everett2}.  More recently,
the attempt to understand how probability
emerges in Everett's approach has focussed on
decision theory (for a review see~\cite{greaves}).
A number of investigators have attempted to derive
the square amplitude measure and Born's rule
from natural decision-theoretic postulates.
Specifically, they try to show that
a `rational' agent who believes he
or she lives in an Everettian universe will make
decisions as if the square amplitude measure
gave chances for outcomes.  The success of this
program, however, is controversial~\cite{kent}.

A necessary condition for encountering distinct
outcomes in Everett's approach is that a product
state evolves into an entangled state, i.e.
\begin{equation}
|\phi\rangle |\psi\rangle \rightarrow
\sum_i c_i |\phi_i\rangle |\psi_i\rangle
= \sum_i c_i' |\phi_i' \rangle |\psi_i' \rangle .
\label{split}
\end{equation}
Although the representation on the left-hand side
is essentially unique, the one on the right-hand side
never is.  Yet outcomes are understood as singling
out a specific representation.  This is the
preferred-basis problem.  In some circumstances
decoherence theory may help to identify the
appropriate basis~\cite{schlosshauer}.

With respect to multiplicity, Everett's approach
has been interpreted as involving (i) many worlds,
(ii) many minds or, more recently, (iii) decohering
sectors of the wave function.  Although the themes
of splitting and multiplicity are undoubtedly
present in Everett's original paper and thesis,
the ontological basis of the approach is never made
entirely explicit in this early work.
There is, however, evidence
that later in his life Everett was thinking in
parallel-universes terms~\cite{deutsch1}.

I have advocated elsewhere~\cite{marchildon1} that
in quantum mechanics, the basic question of
interpretation can be formulated as ``How can the
world be for quantum mechanics to be true?'' This
is related to what has been called the \emph{semantic
view} of theories~\cite{fraassen}.  There can be many
consistent answers to the question just raised, and
each one adds understanding.  To achieve this,
however, each answer should be formulated
in as precise and as clear a manner as possible.

Within such a research program, it seems entirely
acceptable if the probability measure is specified as
an additional postulate, the way Everett did it or
in terms of the frequency operator introduced
by Hartle~\cite{hartle} and Graham.  Likewise
the preferred basis can be associated with
beables corresponding to hidden variables,
perhaps motivated by decoherence.  There is,
however, a pressing need to understand the
true nature of multiplicity.
%
\section{Multiplicity: many worlds and many minds}
\label{sec3}
The idea of a real split into
a multitude of worlds was first made explicit
by DeWitt~\cite{dewitt}.

One of the first questions that come to mind is
whether the split occurs every time that, as
in~(\ref{split}), a product state transforms into
an entangled state, even if the process is
purely microscopic?  DeWitt seems to answer the
question in the affirmative.  Since microscopic
processes can often easily be reversed, the split
itself has to be reversible, or at least reflect that
reversibility.

If the split occurs only in some processes, like
Everett himself seems to have been
thinking~\cite{everett1}, one should specify the
precise conditions (enough mass, enough
particles, $\ldots$)
in which it takes place.  In other words,
one must add elements to minimal quantum mechanics.
Vaidman, for instance, defines the concept of a world
in terms of macroscopic objects, and writes the quantum
state of a world as~\cite{vaidman}
\begin{equation}
|\Psi_{\text{world}} \rangle
= |\Psi\rangle_{\text{object}_1}
|\Psi\rangle_{\text{object}_2} \ldots
|\Psi\rangle_{\text{object}_N} |\Phi \rangle .
\end{equation}
Since object$_i$ is macroscopic, this immediately
raises the question of the classical-quantum distinction.
Further interrogations involve the precise time when
the split occurs, whether it occurs on an equal time
hypersurface or on the light cone, etc.

Instead of splitting in the strict sense,
Deutsch~\cite{deutsch2} postulates a process of
bifurcation.  There is at any time an infinite number
of worlds, which neither increases nor decreases.
A bifurcation at a given time is associated with a
particular \emph{interpretation basis}.

If splitting is restricted to specific processes,
it can be taken as reversible or irreversible.
An irreversible split will entail differences
with unitary quantum mechanics.  Indeed in this
case, the state in a given world is given by
$|\phi_i\rangle |\psi_i\rangle$.  The reverse
measurement interaction would not bring this back
to the initial state (although it would bring
$\sum_i c_i |\phi_i\rangle |\psi_i\rangle$
back to the initial state).

Such are questions that must be answered for
the many-worlds approach to be well-defined.
Answering them is also necessary for the quantum
measurement problem to be solved, since the gist of
the solution precisely consists in the splitting
of worlds.

The many-minds view~\cite{albert} places the split in
consciousness rather than in the outside world.
There is, in the words of Lockwood~\cite{lockwood},
``no good reason for supposing
that the apparent macroscopic definiteness of
the world is anything other than an artefact
of our own subjective point of view.''

In the many-minds approach, kets $|\psi _i\rangle$
in~(\ref{split}) involve brain states and these
are in a quantum superposition.  There can either
be one mind (believing~$i$ with probability
$|c_i|^2$), or an infinite number of minds,
supervening on brain states.

Similar questions can be raised in the many-minds
view as in the many-worlds view.  What kinds of
mind split?  Only human minds, or also cats' minds?
What, in the quantum mechanical formalism, singles out
brain states?  Again these and similar questions must
be answered before anyone can claim to have solved
the measurement problem satisfactorily.
%
\section{Decohering wave function}
\label{sec4}
Everett's way of solving the measurement problem
involves asserting that statements ``Observable
$A$ has value $a_1$'' and ``Observable $A$ has
value $a_2$'' (with $a_1 \neq a_2$) are both
true.  The apparent contradiction is avoided
by construing each statement as ``Observable
$A$ has value $a_i$ relative to value $b_i$
of $B$,'' for $i = 1, 2$.  It has been
argued~\cite{saunders,wallace1} that this
solution of the problem of actuality is
analogous to a solution of the problem of
tense, where apparently contradictory
statements ``Event $E_1$ is now'' and
``Event $E_2$ is now'' are made consistent
by construing them as ``Event $E_i$ is now
relative to event $F_i$,'' for $i = 1, 2$.

Formally, this solution of the problem of tense
consists in adding a dimension to reality.  The
universe is not fully specified through its
spatial, but only through its spatiotemporal,
configuration.  And this, in quantum mechanics,
is essentially what many worlds do.  The universe
(say at a given time) is not adequately specified
by giving a single spatial configuration.  It
involves many configurations, differing in
macroscopic aspects, which can be indexed by an
additional variable that can be viewed as an added
dimension to reality.

The decohering wave function approach to
understanding Everett denies that there is a
genuine split into many worlds or many minds.
The multiplicity is instead associated with
sectors in the decohering universal wave
function, in each of which sectors observables
have values relative to other ones.
To quote Wallace~\cite{wallace2},
``If A and B are to be `live cat' and `dead cat'
then [different micro-world properties] P and Q
[on which A and B supervene] will be described
by statements about the state vector which
(expressed in a position basis) will concern the
wave function's amplitude in vastly separated
regions $R_P$ and $R_Q$ of configuration space, and
there will be no contradiction between these
statements.''

Can this attempt to avoid contradiction be
understood in terms of
added dimensions?  The fact that there is no
genuine split means that the added dimension
referred to earlier, as a label for distinct
worlds, is not available here.  But then it is
not clear why there is no contradiction.
Obviously, the projections in real
three-dimensional space of the live cat
configuration space coordinates cannot overlap
with the projections of, say, the Geiger
counter coordinates.  Why could they overlap
with the projections of the dead cat
coordinates?  Would mass distributions in one
branch behave as `ghosts' to mass distributions in
other branches?

One way to introduce extra dimensions is to use
the ones provided by configuration space.  This
may be what Wallace and others have in mind.
But the problem just mentioned does not disappear.
How are we to make sense of projections in
three-dimensional space, and determine what is
allowed to overlap and what is not?  Moreover,
there is another problem with adding in all the
dimensions of configuration space.  Do we have in
mind the full fine-grained configuration down to
the level of subatomic particles?  Or do we rather
envisage some coarse-grained configuration?
Proponents of the decohering wave function
approach clearly mean the latter.  But then one needs
to specify what level of coarse graining is meant.

That need is denied by Wallace~\cite{wallace2}, who
believes that ``the somewhat blurred borderline between
states where quasi-particles exist and states where they do
not should not undermine the status of the quasi-particles
as real---any more than the absence of a precise point
where a valley stops and a mountain begins should
undermine the status of the mountain as real.''
But the analogy breaks down at a crucial point.
Different levels of coarse graining will never make
two similar mountains stem from just one, whereas
Everettian multiplicity is here understood to appear
at some level of coarse graining only.

To put Everettian multiplicity in perspective,
Allori \emph{et al.}~\cite{allori1} have emphasized
the importance of the concept of \emph{primitive
ontology}, whose elements are the stuff that things
are made of.  Their analysis leads to the
conclusion that they ``do not
see how the existence and behavior of tables and
chairs and the like could be accounted for without
positing a primitive ontology---a description
of matter in space and time.''

Based on that, a specific ontology
for the many-worlds theory, close to Schr\"{o}dinger's
first interpretation of the wave function, was proposed
through a three-dimensional mass density defined
in terms of the full configuration-space
wave function as~\cite{allori2}
\begin{align}
m(\mathbf{x},t) &= \sum_{i=1}^N m_i \int
d\mathbf{x}_1 \ldots d\mathbf{x}_N
\delta (\mathbf{x} - \mathbf{x}_i )
|\Psi (\mathbf{x}_1, \ldots \mathbf{x}_N; t)|^2
\notag \\
&= \sum_{i=1}^N m_i |\psi_i (\mathbf{x}, t)|^2 .
\end{align}
It can be shown that the live cat and the dead cat
both contribute to the total mass density in
three-dimensional space.  In the words of Allori
\emph{et al.}, however, ``the universe according to
[this theory] resembles the situation of a TV set
that is not correctly tuned, so that one always
sees a mixture of several channels.''  Making sense
of this is a challenge that proponents of the
decohering wave function approach have to take up.
%
\section{Conclusion}
\label{sec5}
Everett's approach to the interpretation of
quantum mechanics can be understood in several
different ways.  Some of the questions and
problems left unanswered in Copenhagen quantum
mechanics or in the Dirac and von Neumann collapse
theory also have to be resolved in Everett's
approach.  It nevertheless appears that the more
`extravagant' understanding, namely that of many
worlds, is the one whose basic ontology is
clearest and which provides the logically sharpest
solution to the measurement problem.
%
\section*{Acknowledgements}
I am grateful to the Natural Science and
Engineering Research Council of Canada for
financial support.  It is a pleasure to
thank Ghislain St-Yves and Tony Sudbery
for stimulating discussions.
%

%
\end{document}